\documentclass[fleqn,twoside]{article}
\usepackage{espcrc2}

\usepackage{graphicx}
\usepackage[figuresright]{rotating}

\newcommand{\AmS}{{\protect\the\textfont2
  A\kern-.1667em\lower.5ex\hbox{M}\kern-.125emS}}

\newcommand{\braket}[2]{\langle #1|#2\rangle}

\title{Flavour physics at CLEO-c}

\author{J. Libby\address[OXF]{University of Oxford, Denys Wilkinson Building, Keble Road, 
Oxford, OX1 3RH, United Kingdom} (on behalf of the CLEO collaboration)}

\begin{document}

\begin{abstract}
Three CLEO-c results related to flavour-physics are presented: the determination of the 
strong-phase difference between $D^{0}\rightarrow K^{+}\pi^{-}$ and $D^{0}\rightarrow 
K^{-}\pi^{+}$, $\delta$, the measurement of the coherence parameter and average strong-phase 
difference between $D^{0}\rightarrow K^{+}\pi^{-}\pi^{-}\pi^{+}$ and $D^{0}\rightarrow 
K^{-}\pi^{+}\pi^{+}\pi^{-}$ and measurements of the variation of strong-phase difference between 
$D^{0}$ and $\overline{D^{0}}$ decays to $K^{0}_{S}\pi^{+}\pi^{-}$ over phase space. All 
measurements are important for determining the unitarity triangle angle $\gamma$ from $B\rightarrow 
D^{(*)}K^{(*)}$ decays. Furthermore, the measurement of $\delta$ is important for interpreting 
$D^{0}-\overline{D^{0}}$ mixing. 
\end{abstract}

\maketitle

\section{Introduction}
\label{sec:intro}
The precise determination of the unitarity triangle angle $\gamma$ is a principal goal of flavour 
physics. In particular, measurements of $\gamma$ from tree-level processes, which are insensitive 
to Beyond the Standard Model (BSM) physics, can be compared to measurements in loop processes, 
which are sensitive to virtual corrections from BSM physics. The current average of tree-level 
determinations of $\gamma$ is $(77^{+30}_{-32})^{\circ}$ \cite{bib:CKMfitter}. The most precise 
determinations to date come from measurements of direct $CP$ violation in $B^{\pm}\rightarrow 
DK^{\pm}$, where the $D$ is a $D^{0}$ or $\overline{D^{0}}$ 
decaying to the same final state. Determinations of $\gamma$ from such measurements require 
knowledge of the strong decay parameters of the $D^{0}$ and $\overline{D^{0}}$, in particular the 
strong-phase differences. 

Strong-phase differences and other important parameters of multi-body $D$ decay, relevant to 
$\gamma$ measurements, can be studied in quantum-correlated $D^{0}\overline{D^{0}}$ pairs produced 
in $e^{+}e^{-}$ collisions at a centre-of-mass energy equal to the $\psi(3770)$ mass. The 
charge-conjugation quantum number $(C)$ of the initial and final states is $-1$. Therefore, if one 
$D$ decays into a $CP$ eigenstate such as $\pi^{+}\pi^{-}$ this identifies the other $D$ to be in a 
state of opposite $CP$. The $CP$-tagged rates of decays are sensitive to the strong-phase 
differences required for $\gamma$ measurements. Furthermore, using multi-body $D$ decays to 
determine $\gamma$ requires knowledge of additional parameters related to the presence of 
intermediate resonances; these can also be determined from quantum-correlated $D$-decay data.   

Studies of $D\rightarrow{K^{-}\pi^{+}}$ decays at the $\psi(3770)$ are also sensitive to the  
$D^{0}-\bar{D^{0}}$ mixing parameters as well as the strong-phase difference between doubly Cabbibo 
suppressed (DCS) and Cabbibo favoured (CF) decays $\delta$, which is defined by the relation 
$\braket{K^{+}{\pi}}{D^{0}}/\braket{K^{-}\pi^{+}}{D^{0}}=re^{i\delta}$, where $r=0.0616$ 
\cite{bib:pdg06} is the absolute value of the amplitude ratio.   

These proceedings describe three measurements by CLEO-c of quantum-correlated $D$-decay. 
Section~\ref{sec:cleoc} describes the CLEO-c experiment and data sets used for the analyses. 
Section~\ref{sec:deltaKpi} presents a measurement of charm mixing parameters and $\delta$. 
Sections~\ref{sec:coherence} and \ref{sec:dalitz} describe measurements of strong parameters 
related to $\gamma$ measurements in the multi-body decays $D\to K^{\pm}\pi^{\mp}\pi^{+}\pi^{-}$ and 
$D\to K^{0}_{S}\pi^{+}\pi^{-}$, respectively.

\section{CLEO-c}
\label{sec:cleoc}
All measurements presented are made with $e^{+}e^{-}\to \psi(3770)$ data accumulated at the Cornell 
Electron Storage Ring 
(CESR). The CLEO-c detector is used to collect these data, which is described in detail elsewhere 
\cite{bib:cleoc}. The total integrated luminosity of the data is $818~\mathrm{pb}^{-1}$, however, 
only $218~\mathrm{pb}^{-1}$ have been used so far for the measurement of $\delta$ presented in 
Section~\ref{sec:deltaKpi}.

\section{Measurement of mixing and the strong-phase difference in $D\rightarrow K^{-}\pi^{+}$}
\label{sec:deltaKpi}
Charm mixing is described by two dimensionless parameters $x\equiv (M_{2}-M_{1})/\Gamma$ and 
$y\equiv(\Gamma_{2}-\Gamma_{1})/2\Gamma$, where $M_{1,2}$ and $\Gamma_{1,2}$ are the masses and 
widths, respectively, of the $CP-$ $(D_1)$ and $CP+$ $(D_2)$ neutral $D$ meson eigenstates and 
$\Gamma\equiv(\Gamma_1+\Gamma_2)/2$. Mixing has been probed in lifetime measurements to 
$CP$-eigenstates and in DCS decays.\footnote{For a review of charm mixing results see Ref. 
\cite{bib:bigdog}.} However, when measuring mixing with the DCS decay $D^{0}\rightarrow 
K^{+}\pi^{-}$ the quantities measured are $R_M\equiv(x^{2}+y^{2})/2$ and $y^{\prime}\equiv 
y\cos{\delta}-x\sin{\delta}$. Therefore, without a direct determination of $\delta$ it is 
impossible to combine or compare the determinations of $y$ from other measurements to those from  
$D^{0}\rightarrow K^{+}\pi^{-}$. Furthermore, constraining $\delta$ is an important input to 
measuring $\gamma$ from $B^{\pm}\to D(K^{+}\pi^{-})K^{\pm}$ decays \cite{bib:ads}; this method is 
discussed further in Section~\ref{sec:coherence}.

\begin{table}[t]
\caption{Correlated ($C$-odd) effective $D^{0}\bar{D}^{0}$ branching fractions to leading order in 
conjugate modes are applied. The rates are normalised to the multiple of the uncorrelated branching 
fractions.}
\label{tab:tqcarelations}
\newcommand{\m}{\hphantom{$-$}}
\newcommand{\cc}[1]{\multicolumn{1}{c}{#1}}
\renewcommand{\arraystretch}{1.2} 
\begin{center}
\footnotesize
\begin{tabular}{@{}lc}
\hline
Mode          & Correlated branching fraction  \\ \hline 
$K^{-}\pi^{+}~vs.~K^{-}\pi^{+}$ & $R_{M}$ \\
$K^{-}\pi^{+}~vs.~K^{+}\pi^{-}$ & $(1+R_{WS})^2 - 4 r \cos{\delta}(r\cos{\delta}+y)$ \\
$K^{-}\pi^{+}~vs.~S_{\pm}$ & $1+R_{WS}\pm 2r\cos{\delta}\pm y$ \\
$K^{-}\pi^{+}~vs.~e^{-}$ & $1-ry\cos{\delta}-rx\sin{\delta}$\\  
$S_{\pm}~vs.~S_{\pm}$ & $0$\\  
$S_{+}~vs.~S_{-}$ & $4$\\
$S_{\pm}~vs.~e^{-}$ & $1\pm y$\\ \hline
\end{tabular}
\vspace{-1cm}
\end{center}
\end{table}

The first measurements of the parameters $y$ and $\cos{\delta}$ in quantum-correlated $\psi(3770)$ 
data are presented; the method, which is known as The Quantum Correlated Analysis (TQCA), follows 
that described in Ref.~\cite{bib:tqca_concept}. The parameters are extracted by comparing the decay 
rates where either one (single-tagged) or both (double-tagged) neutral $D$ decays are 
reconstructed. The effective double-tagged rates for different final states are summarised in 
Table~\ref{tab:tqcarelations}, where $S^{\pm}$ and $e^{\pm}$ are used  to indicate $CP\pm$ 
eigenstates and semileptonic final states, respectively; some rates depend on the $D^{0}\to 
K^{-}\pi^{+}$ wrong-sign rate ratio, $R_{WS}\equiv r^{2}+ry^{\prime}+R_{M}$. The most striking 
consequences of the quantum correlations are the enhancements and suppression of the 
$D^{0}\overline{D^{0}}$ decaying to opposite and the same $CP$ eigenstates, respectively. 

\begin{table}[thb]
\caption{Final states reconstructed in TQCA.}
\label{tab:tqcastates}
\newcommand{\m}{\hphantom{$-$}}
\newcommand{\cc}[1]{\multicolumn{1}{c}{#1}}
\renewcommand{\arraystretch}{1.2} 
\begin{center}
\footnotesize
\begin{tabular}{@{}lc}
\hline
Type          & Final states \\ \hline 
Flavoured     & $K^{-}\pi^{+}$, $K^{+}\pi^{-}$ \\
$S^{+}$       & $K^{+}K^{-}$, $\pi^{+}\pi^{-}$, $K^{0}_{S}\pi^{0}\pi^{0}$, $K^{0}_{L}\pi^{0}$ \\
$S^{-}$       & $K^{0}_{S}\pi^{0}$, $K^{0}_{S}\eta$, $K^{0}_{S}\omega$ \\
$e^{\pm}$     & Inclusive $Xe^{+}\nu$, $Xe^{-}$ \\
\hline
\end{tabular}
\end{center}
\vspace{-1cm}
\end{table}

The analysis has been performed with $218~\mathrm{pb}^{-1}$ of $\psi(3770)$ data \cite{bib:tqca}. 
The single and double-tag rates have been determined for the final states listed in 
Table~\ref{tab:tqcastates}. Hadronic final states without a $K^{0}_{L}$ are fully reconstructed via 
two kinematic variables: the beam-constrained candidate mass 
$M\equiv\sqrt{E_{beam}^2-\mathbf{p}_{D}^{2}}$ and $\Delta E\equiv E_{D}-E_{beam}$, where $E_{beam}$ 
is the beam energy, $\mathbf{p}_{D}$ and $E_{D}$ are the $D^{0}$ candidate momentum and energy, 
respectively. The yields are extracted from the one or two dimensional $M$ distributions for single 
and double-tagged events, respectively. The reconstruction of $K^{0}_{L}\pi^{0}$ events utilises 
the missing-mass technique described in Ref.~\cite{bib:K0Lpi0}. The inclusive $e^{\pm}$ tags 
exploit electrons identified using a multivariate discriminant \cite{bib:elecid}.

\begin{table}[t]
\caption{TQCA results from the standard and extended fit. Uncertainties are statistical and 
systematic, respectively.}
\label{tab:tqcaresults}
\newcommand{\m}{\hphantom{$-$}}
\newcommand{\cc}[1]{\multicolumn{1}{c}{#1}}
\renewcommand{\arraystretch}{1.2} 
\begin{center}
\footnotesize
\begin{tabular}{@{}lcc}
\hline
Parameter          & Standard fit & Extended fit  \\ \hline 
$y~(10^{-3})$ & $-45\pm 59 \pm 15$ & $6.5\pm 0.2 \pm 0.1$ \\
$r^{2}~(10^{-3})$ & $8.0 \pm 6.8 \pm 1.9$ & $3.44 \pm 0.01 \pm 0.09$ \\
$\cos{\delta}$ & $1.03\pm 0.19 \pm 0.06$ & $1.10\pm 0.35 \pm 0.07$ \\
$x^{2}~(10^{-3})$ & $-1.5\pm 3.6 \pm 4.2$ & $0.06 \pm 0.01 \pm 0.05$ \\
$x\sin{\delta}~(10^{-3})$ & 0 (fixed) & $4.4 \pm 2.4 \pm 2.9$ \\
\hline 
\end{tabular}
\end{center}
\vspace{-1cm}
\end{table}

\begin{figure}[htb]
\includegraphics*[width=1.0\columnwidth]{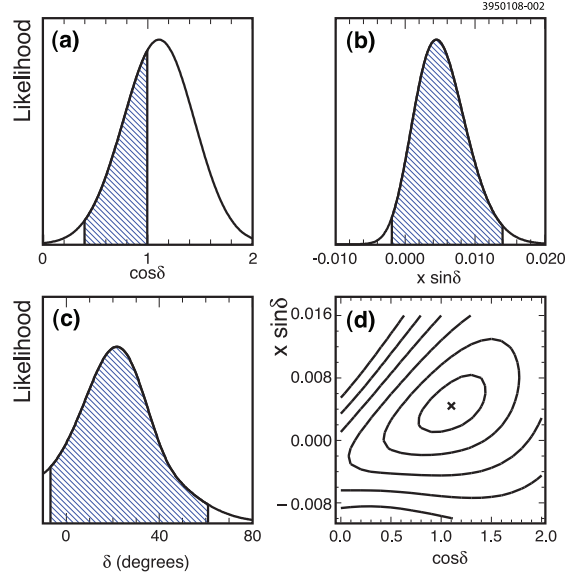}
\vspace{-1cm}
\caption{Extended fit likelihood including both statistical and systematic uncertainties for 
$\cos{\delta}$ (a), $x\sin{\delta}$ (b), $\delta$ (c) and simultaneous likelihood for 
$\cos{\delta}$ and $x\sin{\delta}$ (d) shown as contours in increments of $1\sigma$. The hatched 
regions contain $95\%$ of the area in the physical regions.}
\label{fig:tqcaresults}
\end{figure}

The standard fit to determine the mixing parameters and $\cos{\delta}$ includes independent, 
external measurements of $R_{M}$, $R_{WS}$ and uncorrelated branching fractions. Additional 
external measurements of charm mixing are included in an extended fit which allows $x\sin{\delta}$ 
to also be determined. The fit accounts for correlations amongst the inputs and any cross-feed 
between the signal channels. 

The results of the standard and extended fit are given in Table~\ref{tab:tqcaresults}. A value of 
$x\sin{\delta}$ can only be determined reliably in the extended fit. The extended fit likelihoods 
for $\cos{\delta}$, $x\sin{\delta}$ and $\delta$ are shown in Figure~\ref{fig:tqcaresults}. Despite 
the additional information in the extended fit the uncertainty on $\cos{\delta}$ increases due to 
the non-linear relationship between $\cos{\delta}$ and $y$. However, the determination of 
$x\sin{\delta}$ in the extended fit allows the likelihood for $\delta$ to be determined. This leads 
to a measurement of $\delta=(22^{+11+9}_{-12-11})^{\circ}$, which is the first direct 
determination. 
These measurements are important for the interpretation of the combined results on charm mixing 
\cite{bib:schwartz_hfag}.

\section{Measurement of the coherence parameter and average strong-phase difference in 
$D\rightarrow K\pi\pi\pi$}
\label{sec:coherence}
The amplitudes for $B^{-}\to D^{0}K^{-}$ and $B^{-}\to \overline{D^{0}}K^{-}$ have a relative phase 
between them of $\delta_B-\gamma$, where $\delta_B$ is the strong-phase difference. For decays in 
which the $D^{0}$ and $\overline{D^{0}}$ decay to the same final state, $f_D$, the two amplitudes 
interfere giving sensitive to $\gamma$. The amplitude for $B^{-}\to D(f_D)K^{-}$ is given by:
\begin{equation}
 A(B^{-}\to D(f_D)K^{-})\propto A_{D^{0}}+r_B e^{i(\delta_B-\gamma)} A_{\overline{D^{0}}} \; ,
\end{equation}
where $r_B<0.13$ at 90\% c.l. \cite{bib:CKMfitter} is the ratio of the absolute value of $B^{-}\to 
\overline{D^{0}}K^{-}$ to $B^{-}\to D^{0}K^{-}$ amplitudes, and $A_{D^{0}}$ and 
$A_{\overline{D^{0}}}$ are the amplitudes for $D^{0}\to f_D$ and $\overline{D^{0}}\to f_D$, 
respectively. 

For the case where $f_D$ is $K^{-}\pi^{+}$ \cite{bib:ads} the decay rates are equal to:
\begin{eqnarray*}
 \Gamma_{B^{-}\to D(K^{-}\pi^{+})K^{-}}  & \propto & 1 + (r_{B}r)^2 \nonumber \\
   & + & 2r_{B}r\cos{(\delta_{B}-\delta-\gamma)} \; , \label{eqn:rsads}\\
	 \Gamma_{B^{-}\to D(K^{+}\pi^{-})K^{-}} & \propto &  r_{B}^{2} + r^2 \nonumber \\ 
	  & + & 2r_{B}r\cos{(\delta_{B}+\delta-\gamma)} \; \label{eqn:wsads}.
\end{eqnarray*}   
The suppressed rate has an interference term of the same order as the other terms leading to 
enhanced sensitivity to $\gamma$.

The rates for $f_D=K^{-}\pi^{+}\pi^{+}\pi^{-}$ are of a similar form, however, an additional 
coherence parameter, $R_{K3\pi}$, is introduced \cite{bib:coherence_theory} to account for the 
possibility of several different intermediate states, with differing strong phases, contributing, 
such as $K^{*0}\rho^{0}$ and $K(1270)^{-}_{1}\pi^{+}$. For example the suppressed rate is given by:
\begin{displaymath}
	\Gamma_{B^{-}\to D(K^{+}\pi^{-}\pi^{-}\pi^{+})K^{-}} \propto r_{B}^{2} + (r_{D}^{K3\pi})^2
\end{displaymath}	 
\begin{displaymath}
 + 2r_{B}r_{D}^{K3\pi}R_{K3\pi}\cos{(\delta_{B}+\delta^{K3\pi}_D-\gamma)} \; ,
\end{displaymath} 
where $r_{D}^{K3\pi}$ is the absolute of the ratio of DCS to CF amplitudes and $\delta^{K3\pi}_D$ 
is the average strong-phase difference. 

The definition of $R_{K3\pi}$ is such that it takes a value between zero and one, with values 
approaching zero corresponding to many different intermediate states contributing equally and 
values approaching one corresponding to the dominance of a single intermediate final state. 
To exploit fully the sensitivity to $\gamma$, $\delta_B$ and $r_B$ in the case of $D\to 
K^{-}\pi^{+}\pi^{+}\pi^{-}$ external measurements of the coherence parameter are extremely 
important.  

The quantum correlated $D^{0}\overline{D^{0}}$ data allow direct measurements of $R_{K3\pi}$ and 
$\delta_D^{K3\pi}$. Double-tagged rates are used to determine these parameters; their dependence on 
the parameters are listed in Table~\ref{tab:coherenceresults}. The interpretation of the 
measurement of $K^{\pm}\pi^{\mp}\pi^{-}\pi^{+}~vs.~K^{\pm}\pi^{-}$ requires the meaurement of 
$\delta$ presented in Section~\ref{sec:deltaKpi}.

\begin{table*}[t]
\caption{Double-tagged rates of interest and their dependence $R_{K3\pi}$ and $\delta_{D}^{K3\pi}$. 
The background subtracted yields from the $818~\mathrm{pb}^{1}$ are shown along with the 
corresponding result for each measurement. Uncertainties are statistical and systematic, 
respectively.}
\label{tab:coherenceresults}
\newcommand{\m}{\hphantom{$-$}}
\newcommand{\cc}[1]{\multicolumn{1}{c}{#1}}
\renewcommand{\arraystretch}{1.2} 
\begin{center}
\footnotesize
\begin{tabular}{@{}lcc}
\hline
$K^{\pm}\pi^{\mp}\pi^{+}\pi^{-}$ vs. & Measurement & Signal yield \\ \hline 
$K^{\pm}\pi^{\mp}\pi^{+}\pi^{-}$ & $R_{K3\pi}^{2}= 0.00\pm 0.16 \pm 0.07$ & $30\pm 6$ \\ 
$CP$-tags & $R_{K3\pi}\cos{\delta^{K3\pi}_D} = -0.60\pm 0.19 \pm 0.24$ & $2183\pm 47$ \\
$K^{\pm} \pi^{\mp}$ & $R_{K3\pi}\cos{(\delta-\delta^{K3\pi}_D)} = -0.20\pm 0.23 \pm 0.09$ & $38\pm 
6$ \\
\hline 
\end{tabular}
\end{center}
\end{table*}

The analysis has been performed on $818~\mathrm{pb}^{-1}$ of data and proceeds in a similar manner 
to that presented in Section~\ref{sec:deltaKpi}.
All the $CP$-tags listed in Table~\ref{tab:tqcastates} are used in this analysis apart from 
$K^{0}_{L}\pi^{0}$; this and other $K^{0}_{L}$ modes will be included in an updated version of the 
analysis. In addition, the $CP$-tags $K^{0}_{S}\phi$ and $K^{0}_{S}\eta^{\prime}$ are included. The 
backgrounds levels are between 1 and 7\% depending on the final state reconstructed. The efficiency 
for reconstruction for the different final states varies between 4 and 30\%.  

\begin{figure}[htb]
\includegraphics*[width=1.0\columnwidth]{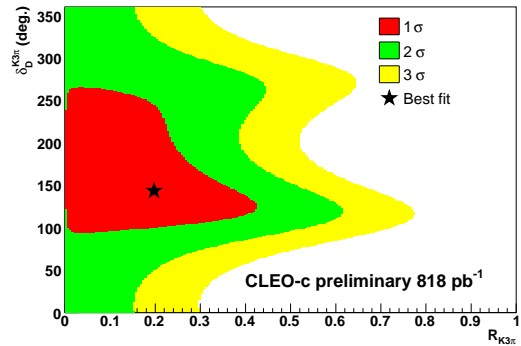}
\vspace{-1.4cm}
\caption{The limits on $R_{K3\pi}$ and $\delta_{D}^{K3\pi}$ at the 1, 2 and $3\sigma$ level.}
\vspace{-0.6cm}
\label{fig:coherenceresults}
\end{figure}

The different constraints on $R_{K3\pi}$ and $\delta_D^{K3\pi}$ determined from the data are given 
in Table~\ref{tab:coherenceresults}. The combined result for the nine $CP$ tagged rates is shown; 
the combination is performed including systematic uncertainties and accounting for all correlations 
amongst the measurements. The combination of the three different constraints leads to the 
likelihood contours in $R_{K3\pi}$ and $\delta_D^{K3\pi}$ space shown in 
Figure~\ref{fig:coherenceresults}. The most likely values of the parameters are $R_{K3\pi}=0.2$ and 
$\delta_D^{K3\pi}=144^{\circ}$. A low value of the coherence parameter is favoured. Low coherence 
dilutes the sensitivity to $\gamma$ of $B\to D(K^{\pm}\pi^{\mp}\pi^{-}\pi^{+})K$ decays, however, 
the measurements are very sensitive to $r_B$ which is a common parameter with other $B\to DK$ 
decays. The usefulness of these constraints on $R_{K3\pi}$ and $\delta_D^{K3\pi}$ can be 
illustrated when they are combined with the expected yields for $B\to D(K\pi) K$ and $B\to 
D(K3\pi)K$ at the LHCb experiment \cite{bib:lhcb,bib:mitesh_dc04,bib:frank_dc04} to determine 
$\gamma$. Depending on the parameter values the uncertainty on $\gamma$ reduces by 25 to 35\%. More 
details of this analysis are documented in Ref.~\cite{bib:frank_llwi}. 

\section{Studies of the strong-phase variation in $D\rightarrow K^{0}_{S}\pi^{+}\pi^{-}$}
\label{sec:dalitz}
\begin{figure}[htb]
\includegraphics*[width=1.0\columnwidth]{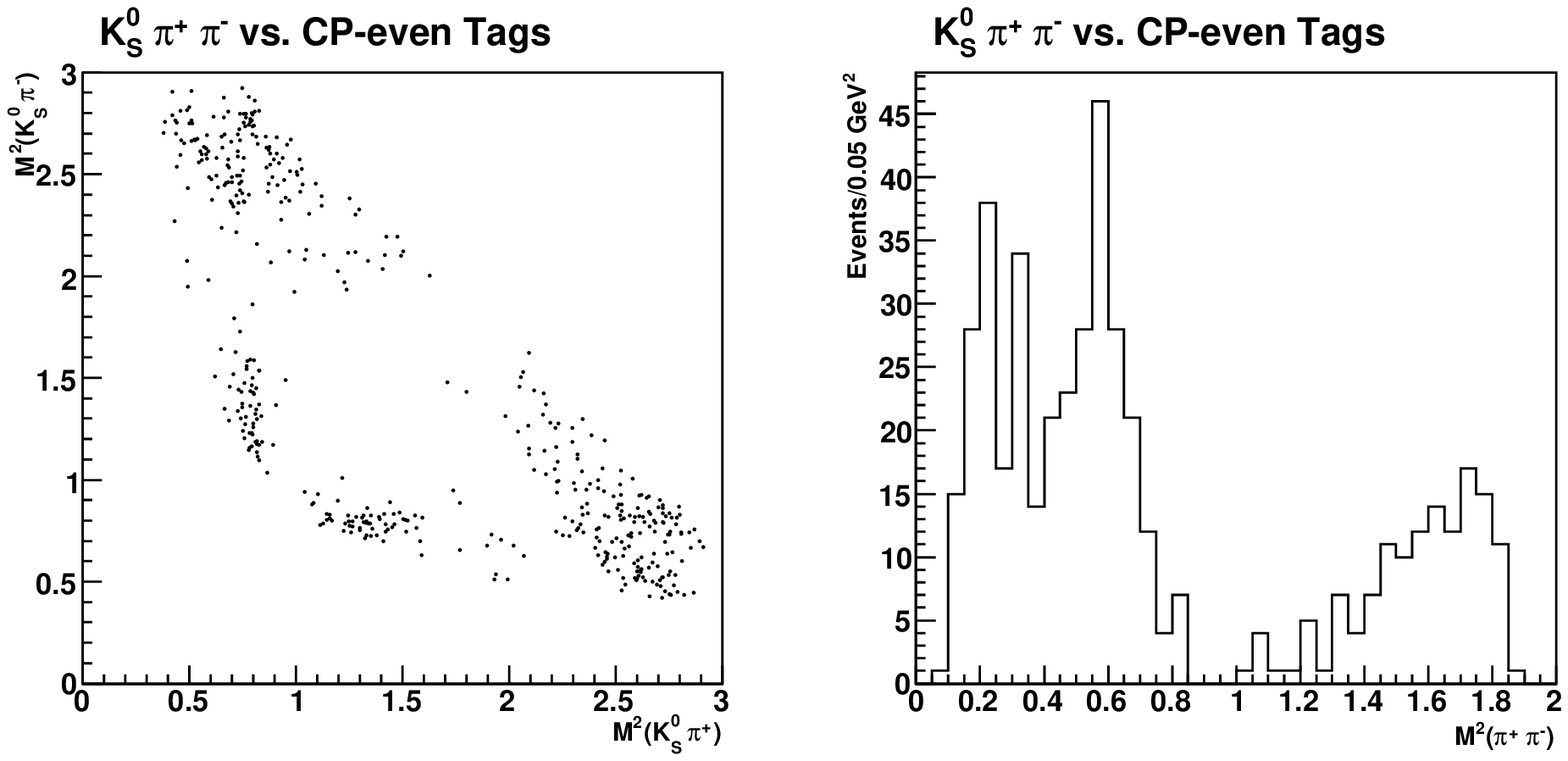}\\
\includegraphics*[width=1.0\columnwidth]{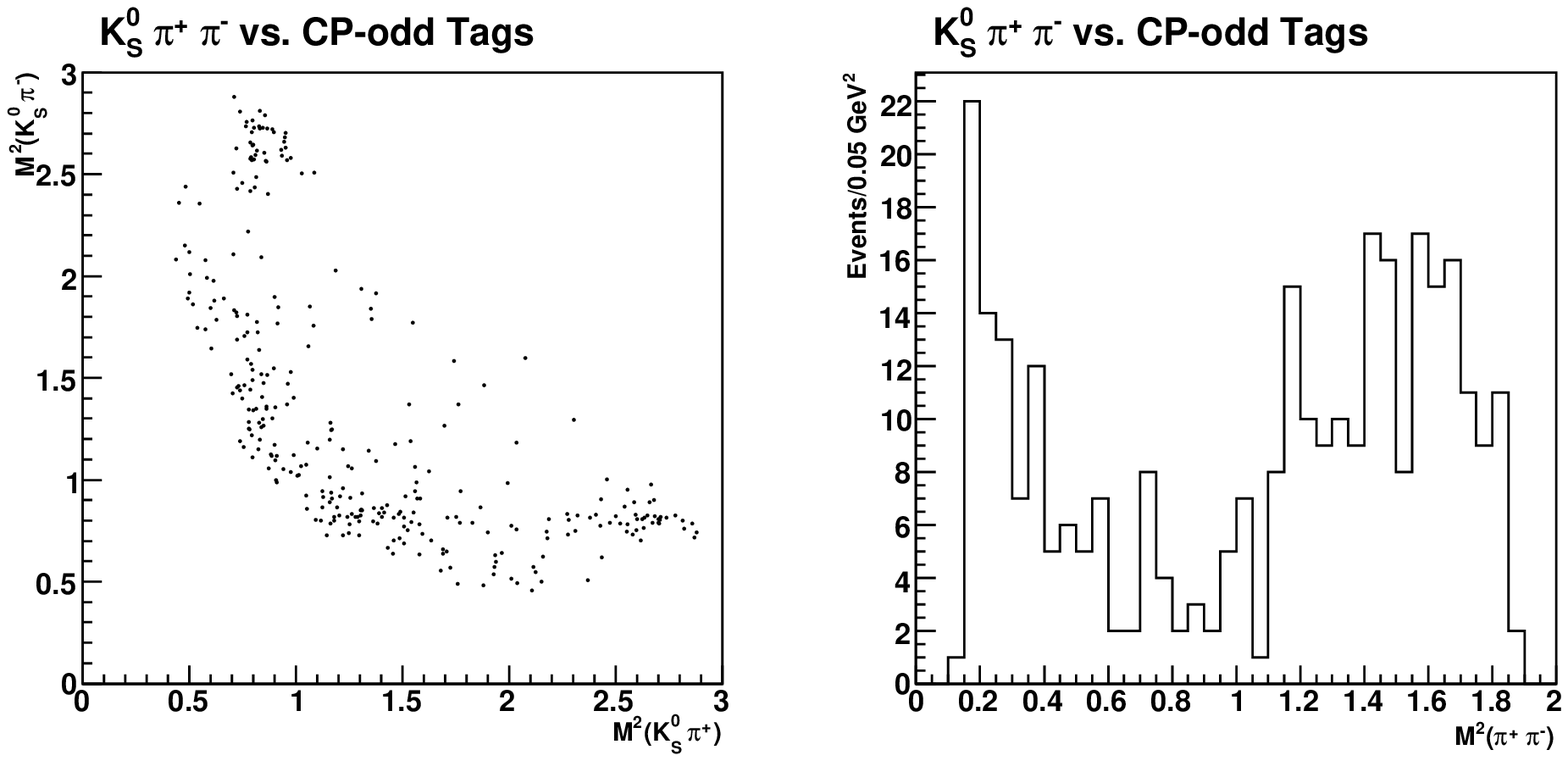}\\
\vspace{-1.5cm}
\caption{The distributions of the $K^{0}_S\pi^{+}$ invariant-mass squared vs. $K^{0}_S\pi^{-}$ 
invariant-mass squared (left) and $\pi^{+}\pi^{-}$ invariant-mass squared for $CP$-even tagged 
(upper) and $CP$-odd tagged (lower) $K^{0}_{S}\pi^{+}\pi^{-}$ events. The data sample corresponds 
to a luminosity of $818~\mathrm{pb}^{-1}$.}
\label{fig:dalitz}
\end{figure}

The process $B^{\pm}\to D(K^{0}_S\pi^{+}\pi^{-})K^{\pm}$ currently provides the best determinations 
of $\gamma$ \cite{bib:babar08,bib:belle08}. The sensitivity to $\gamma$ is exploited via likelihood 
fits to the $K^{0}_S\pi^{+}\pi^{-}$ Dalitz plot \cite{bib:ggsz}. These fits require models of the 
$D^{0}\to K^{0}_{S}\pi^{+}\pi^{-}$ amplitudes which introduce a $5^{\circ}$ to $10^{\circ}$ 
systematic uncertainty on $\gamma$. Future measurements of $\gamma$ will be limited by this 
uncertainty. 

An alternative method is to perform a model-independent binned analysis of the Dalitz plots, which 
was orginally proposed in Ref.~\cite{bib:ggsz} and has been developed significantly in 
Ref.~\cite{bib:bnp}. The binned method requires knowledge of two parameters in each bin of the 
Dalitz space, $c_i$ and $s_i$, which are the average cosine and sine of the strong-phase difference 
between 
$D^{0}$ and $\overline{D^{0}}$ decaying to $K^{0}_S\pi^{+}\pi^{-}$, respectively. 

The values of $c_i$ and $s_i$ can be determined from $\psi(3770)$ data at CLEO-c \cite{bib:eric}. 
The value of $c_i$ can be measured from differences in the $CP+$ and $CP-$ tagged Dalitz plots; 
$s_i$ and $c_i$ are determined from double $K^{0}_{S}\pi^{+}\pi^{-}~vs.~K^{0}_{S}\pi^{+}\pi^{-}$ 
events. 

The measurement of $c_i$ and $s_i$ from CLEO-c's  $818~\mathrm{pb}^{-1}$ sample of $\psi(3770)$ 
data is underway. Again $M$ and $\Delta E$ are the main variables used to isolate the signal. The 
$CP$-tagged Dalitz plots and $\pi^{+}\pi^{-}$ invariant-mass squared distributions are shown in 
Figure~\ref{fig:dalitz}. The presence of the quantum correlations is shown clearly by the absence 
of the $\rho^{0}$ in the $CP-$ tagged data.

Preliminary estimates indicate that these data will lead to a 3 to $5^{\circ}$ uncertainty on 
$\gamma$ from those on the measurements of $c_i$ and $s_i$ at CLEO-c. The measurements can be 
improved by including information from $K^{0}_{L}\pi^{+}\pi^{-}$ decays at CLEO-c.

\end{document}